\documentclass[aps,prL,twocolumn,superscriptaddress,showpacs]{revtex4}
\usepackage[colorlinks=true,urlcolor=blue,citecolor=blue,linkcolor=blue]{hyperref}
\usepackage{graphicx}
\usepackage{latexsym}
\usepackage{stmaryrd}
\usepackage{amsmath}
\usepackage{amssymb}
\usepackage{epstopdf}
\DeclareMathOperator{\diag}{diag}

\begin{document}
\makeatletter
\newcommand{\rmnum}[1]{\romannumeral #1}
\newcommand{\Rmnum}[1]{\expandafter\@slowromancap\romannumeral #1@}
\newcommand{\B}[1]{{\textcolor{blue}{#1}}}
\makeatother
	
\title{Quantized charge-pumping in higher-order topological insulators}

\author{Bing-Lan Wu }
\affiliation{Gusu Laboratory of Materials, Suzhou 215123, China}
\affiliation{Institute for Advanced Study, Soochow University, Suzhou 215006, China}
\author{Ai-Min Guo}\email{aimin.guo@csu.edu.cn}
\affiliation{Hunan Key Laboratory for Super-microstructure and Ultrafast Process, School of Physics and Electronics,Central South University, Changsha 410083, China}
\author{Zhi-Qiang Zhang }\email{zhangzhiqiangphy@163.com}
\affiliation{School of Physical Science and Technology, Soochow University, Suzhou, 215006, China}
\affiliation{Institute for Advanced Study, Soochow University, Suzhou 215006, China}
\author{Hua Jiang}
\affiliation{School of Physical Science and Technology, Soochow University, Suzhou, 215006, China}
\affiliation{Institute for Advanced Study, Soochow University, Suzhou 215006, China}
\date{\today}
	
\begin{abstract}
We study the quantized charge pumping of higher-order topological insulators (HOTIs) with edge-corner correspondences based on the combination of the rotation of in-plane magnetic field and the quantum spin Hall effect. A picture of a specific charge pumping process is uncovered with the help of the non-equilibrium Green's function method.
Significantly, we demonstrate that the quantized charge pumping current is achieved without the participation of bulk states, and the charges move along the boundary of the sample.
Furthermore, the effects of external parameters on the pumping current is also studied. We find that the magnitude and direction of the pumping current can be manipulated by adjusting the coupling strength between the leads and sample. Our work deepens the understanding of the charge pumping in HOTIs and extends the study of their transport properties.
	
\end{abstract}
	
\maketitle

\section{Introduction}\label{section1}

As the analog of classical pumping, quantum charge pumping is one of the most important transport phenomena in condensed matter physics \cite{01,02,03,04,05,06,07,08,09,10,11,12,13,14,15,16,17,18,19,20,20a1}.
The combination of topology and charge pumping has attracted significant interest over the past decades \cite{04,05,06,07,08,09,10,11,12,13,14,15,16,17,18,19,20}.
In the study of the quantum Hall \cite{21,22} and quantum spin Hall effects \cite{23,24,25}, the pumping process can be used to identify the quantization of their topological invariants and suggests the existence of extended edge states in the bulk gap.
For the one-dimensional samples, the topological order can always be identified with the help of quantized charge pumping \cite{20,27}.
These studies strongly suggest that the quantized charge pumping is closely related to the intrinsic topological features.

Generally, topological insulators \cite{21,22,23,24,25} exhibit the bulk-boundary correspondence, where the quantized properties can be characterized by bulk states \cite{21}.
Therefore, the bulk states always directly participate in the charge pumping process to demonstrate their topological natures.
 Recently, the concept of higher-order topological insulators (HOTIs)
 \cite{30,31,32,33,34,35,36,37,HOTI0,HOTI1,HOTI2,HOTI3,HOTI4,HOTI5,HOTI6,HOTI7,
 HOTI8,HOTI9,HOTI10,HOTI11,HOTI12,HOTI13,HOTI14,38,39,HOTIpump1,HOTIpump2,HOTIpump3} has been put forward, which goes beyond the conventional bulk-boundary correspondence.
 Specifically, the $ n$-${th}$-order HOTI in $ d $-dimension captures the topological states in its ($ d-n $)-dimensional boundaries \cite{30,31,32} with $n>1$.
For two-dimensional HOTIs, their bulk and edge states are both gapped.
However, the zero-energy states locate at the corners of the samples \cite{HOTI14,38,39}. Notably, one of the specific features of HOTIs is that the conventional bulk-boundary correspondence could collapse \cite{HOTI14}, and the corner states characterizing the unconventional topological order of HOTIs could move along the edge of the sample by only manipulating the edge states' mass domain walls \cite{38}. Therefore, a quantized charge pumping without the participation of bulk states may exist. Although the charge pumping in HOTIs has been studied very recently \cite{HOTIpump1,HOTIpump2}, the combination of charge pumping and HOTIs associated with such unique features is rarely reported.

\begin{figure}[t]
	\centering
	\includegraphics[width=0.44\textwidth]{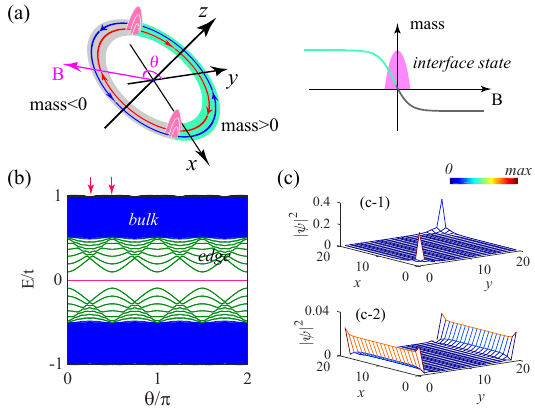}
	\caption{(a) Schematic diagram of the mass domains for HOTIs. The domain wall ensures the bound states of HOTIs as shown in the right figure. The pink arrows shows the direction of the in-plane magnetic filed. (b) The eigenvalues versus angle $ \theta=2\pi f\tau$ with sample size $N \times N=20 \times 20$. $ f $ is the rotational frequency and $\tau $ denotes the time. (c) zero-modes' eigenstates for $\theta$ marked in (b) with upper: $\theta=\pi/4$; lower: $\theta=\pi/2$. }
	\label{f1}
\end{figure}

In this work, we propose the charge pumping process by manipulating the mass domain walls sitting at the edge of two-dimensional HOTIs [see Fig. \ref{f1}(a)], which is closely related to the unique features of HOTIs.
We uncover that the major characteristic of such a pumping process is that the electrons will be transferred from one corner to its opposite one along the edge instead of through the bulk of the sample.
Moreover, the dependence of the external parameters on the pumping current is studied.
We find both the  magnitude and the direction of the pumping current can be efficiently manipulated.
In particular, by adjusting the coupling strength between the leads and sample or the leads' voltages, the current reversal is available.
	
The rest of this paper is organized as follows: In Sec. \ref{section2}, we present the model and the methods. In Sec. \ref{section3} and Sec. \ref{section4}, we demonstrate the charge pumping process in HOTIs. In Sec. \ref{section5}, the manipulation of the pumping currents is studied. Finally, a brief discussion and summary are presented in Sec. \ref{section6}.
	
\section{model and method}\label{section2}

\subsection{model}

We consider a modified Bernevig-Hughes-Zhang model with Hamiltonian \cite{38}:
\begin{align}
	\begin{split}
		H_{c}=\sum_{i}\{T_0c_{i}^{\dagger}c_{i}+[T_{x}c_{i+\delta_{x}}^{\dagger}c_{i}+T_{y}c_{i+\delta_{y}}^{\dagger}c_{i} + h.c.]\},
\label{EQ1}
	\end{split}
\end{align}
where $T_0=-Bsin(\theta)\tau_{0}\sigma_{x}+Bcos(\theta)\tau_{0}\sigma_{y}+m \tau_{z}\sigma_{0}$ and $ T_{x/y}=\frac{t}{2} \tau_{z}\sigma_{0} + \frac{\lambda}{2i} \tau_{x}\sigma_{x/y} $. $ \sigma_{x,y,z} $ ($ \tau_{x,y,z} $) are Pauli matrices in spin (orbital) spaces. $ \sigma_{0} $ ($ \tau_{0} $) is the $ 2\times 2 $ identity matrix. $ B $ is in-plane magnetic field. $ \theta=2\pi f\tau$ denotes the orientation of $ B $ varying from $ 0 $ to $ 2\pi $. $ f $ is the rotational frequency and $\tau $ denotes the time. Parameters are fixed at $m=t$, $\lambda=t$ and $B=0.5t$ throughout the paper. $t$ is the energy unit with $t=1$.

Due to the in-plane magnetic field $B$, mass domain walls sitting at the boundary of the sample appear \cite{38}.
As shown in Fig. \ref{f1}(a), a sample with disk geometry possesses the gapless helical edge states when $B=0$.
After considering the in-plane magnetic field $B$, the interaction between opposite spins leads to the bandgap for the helical edge states.
To be concrete, $ mass $ terms are introduced into the effective model for the helical edge states.
Further, the group velocity of helical edge states changes its sign along the direction of $B$, which reverses the sign of the mass.
The gapped helical edge state with opposite sign of $ mass $ [$ mass>0 $ in the green region and $ mass<0 $ in the gray area, see Fig. \ref{f1}(a)] ensures the existence of the corner states characterizing the topological natures of HOTIs \cite{38}.

For simplicity, we consider a square sample, and its eigenvalues versus $ \theta $ are shown in Fig. \ref{f1}(b).
By checking their zero-energy wavefunction distributions [see Fig. \ref{f1}(c)], one observes that the corner states emerge in a certain angle $ \theta $.
%These results show that the mass domain walls of HOTIs rotate with the rotation of the magnetic field.
These results indicate that the bound states protected by the mass domain walls rotate with the in-plane magnetic field, as shown in Fig. \ref{f1}(c). Thus, the charges carried by the bound states rotate by varying $\theta$, which gives rise to the study of charge pumping in HOTIs without the participation of bulk states.

\subsection{method}

The charge pumping current is calculated by
employing the non-equilibrium Green's function method \cite{40,41}. Taking a square sample with size $ N_{x}=N_{y}=N $ as an example, Eq. (\ref{EQ1}) is rewritten as:
%(2)
\begin{align}
	\begin{split}
		H_{c}&=\sum_{n,m=1}^N T_{0} c_{n,m}^{\dagger}c_{n,m} + \sum_{m=1}^N \sum_{n=1}^{N-1} T_{x}c_{n+1,m}^{\dagger}c_{n,m}\\
		&+ \sum_{n=1}^N \sum_{m=1}^{N-1} T_{y}c_{n,m+1}^{\dagger}c_{n,m} + h.c.
\label{EQ2}
	\end{split}
\end{align}
 Its matrix form ${\mathcal{H}}_c$ can be written as $H_c=\hat{\mathcal{X}}^\dagger {\mathcal{H}}_c\hat{\mathcal{X}}$ with the basis $\hat{\mathcal{X}}=[c_{1,1;\alpha}, c_{1,2;\alpha},\cdots,c_{n,m;\alpha},\cdots,c_{N,N;\alpha}]^T$. $\alpha$ stands for the four orbital freedoms of $T_{0/x/y}$. Supposing the eigenequation is $ {\mathcal{H}}_c\psi_j=E_j\psi_j$, one has:
 %(3)
\begin{align}
	\begin{split}
	 {\mathcal{H}}_c \mathcal{S}=\mathcal{S} \diag[E_{1},E_2\cdots E_{4N^2}].
	\end{split}
\end{align}
 $\mathcal{S}=[\psi_1,\psi_2,\cdots,\psi_{4N^2}]$ is constructed by the eigenvectors $\psi_j$.
 The coefficient $4$ originates from the size of $T_{0/x/y}$.
  Considering $\hat{\mathcal{X}}=\mathcal{S}\hat{\mathcal{K}}$ with basis $\hat{\mathcal{K}}=[b_1,b_2, \cdots,b_{4N^2}]^T$, one has
  %(4)
\begin{align}
	\begin{split}
H_c=\hat{\mathcal{X}}^\dagger {\mathcal{H}}_c\hat{\mathcal{X}}= \hat{\mathcal{K}}^\dagger \mathcal{S}^\dagger  {\mathcal{H}}_c \mathcal{S} \hat{\mathcal{K}}=\sum\limits_{j=1}^{4N^2}E_jb_j^\dagger b_j.
\label{EQ4}
	\end{split}
\end{align}
Here, $ E_{j} $ is the eigenvalue of the $ j$-$th$ eigenstate. Thus, the following relation is ensured for each basis:
%(5)
\begin{equation}
c_{n,m;\alpha}=\sum_{j=1}^{4N^2} \psi_j(n,m; \alpha) b_{j}.
\end{equation}
 $\psi_j(n,m; \alpha)$ is the component of the wavefunction of the $ j$-$th$ eigenvector at site $ (n,m;\alpha) $.

 The Hamiltonian of the leads and their couplings to the sample can be expressed as \cite{40}:
 %(6)
\begin{align}
	\begin{split}
		H_{el}=\sum_{\beta,k} \{\varepsilon_{\beta,k} a_{\beta,k}^{\dagger}a_{\beta,k} + \sum_{q} t_{\beta}[ a_{\beta,k}^{\dagger}c_{q} + h.c.]\}.
	\end{split}
\end{align}
$q=(1,1)\equiv\sum\limits_{\alpha}(1,1;\alpha)$ and $(N,N)\equiv\sum\limits_{\alpha}(N,N;\alpha)$ for the left and right leads, respectively. $ \beta = L/R $ stands for the left/right lead. After considering Eq. (\ref{EQ4}), the above equation can be rewritten as:
%(7)
\begin{align}
	\begin{split}
 \mathcal{H}_{el}=\sum_{\beta,k} \{\varepsilon_{\beta,k} a_{\beta,k}^{\dagger}a_{\beta,k} + \sum_{j=1}^{4N^{2}} t_{\beta}[\psi_{j}(o_{\beta}) a_{\beta,k}^{\dagger} b_{j} + h.c.]\}.
	\end{split}
\end{align}
For simplicity, we set
$ o_{L}=(1,1) $ and $ o_{R}=(N,N) $, which determines the coupling between the left/right lead and the sample's corresponding sites, as shown in Fig. \ref{f3}(b).
%Such a choice also prevents the resonant tunneling between different leads under different $\theta$.   The orbit and spin degrees of freedoms for one site are summarized here after.

Providing that the occupation number \cite{40} of electron for the $ j $-$th$ energy level is $ n_{j}= \langle b_{j}^{\dagger}b_{j}\rangle$, one has
%(8)
\begin{align}
	\begin{split}
		\dfrac{dn_{j}}{dt} = \dfrac{1}{i\hbar}\langle [b_{j}^{\dagger}b_{j},\mathcal{H}_c+\mathcal{H}_{el}]\rangle.
	\end{split}
 \end{align}
The total occupation number $n_{j}$ and the components for the leads $n_{j,L/R}$ under the adiabatic approximation satisfy the following forms \cite{41}:
%(9)
\begin{align}
	\begin{split}
		&\dfrac{dn_{j}}{d\tau} = \dfrac{1}{\hbar} (\widetilde{\Gamma}_{L}f_{L} + \widetilde{\Gamma}_{R}f_{R}) - \dfrac{1}{\hbar}(\widetilde{\Gamma}_{L}+\widetilde{\Gamma}_{R})n_{j},\\
        &\dfrac{dn_{j,L/R}}{d\tau}=\dfrac{1}{\hbar}\widetilde{\Gamma}_{L/R}[f_{L/R}(E_{j}) - n_{j}],
\label{EQ7}
	\end{split}
\end{align}
where $ \widetilde{\Gamma}_{L}= \Gamma_{L}|\psi_{j}(1,1)|^{2} $ and $ \widetilde{\Gamma}_{R} = \Gamma_{R}|\psi_{j}(N,N)|^{2} $. $\hbar$ is the reduced Planck constant.
For one-dimensional metallic leads, the related linewidth function can be set as $ \Gamma_{L/R}=2\pi \rho |t_{L/R}|^{2}$ with the constant $ \rho $ the density of states for the leads \cite{40}, which represents the coupling between the leads and the sample.
Since the similarity transformation $\mathcal{S}$ leads to $ {\Gamma}_{L/R}\rightarrow \widetilde{\Gamma}_{L/R}$, the density of states for the sample play key roles, among which $ |\psi_{j}(1,1)|^{2} $ and $ |\psi_{j}(N,N)|^{2} $ are the electron density of the $j$-$th$ eigenvectors for sites $(n,m)=(1,1)$ and $(N,N)$, respectively.
$ f_{L/R}=[1+e^{\frac{E_F+V_{L/R}}{k_BT_0}}]^{-1} $ is the Fermi-Dirac distribution function. $ V_{L/R} $ is an additional voltage potential on the left/right lead, $ k_{B} $ is the Boltzmann constant. $ T_0 $ is the temperature and is fixed to $T_0\rightarrow0$.

Generally, the current for the $j$-$th$ eigenvalue can be represented as
\begin{equation}
I_{j,L/R}=ef\oint dn_{j,L/R},
\end{equation}
with $I_{L/R}=\sum_{j=1}^{4N^2}I_{j,L/R}$, which is a function of $n_j$. The integration is over one period. The occupation number $n_j$ can be obtained by solving Eq. (\ref{EQ7}) self-consistently in one periodic, and the current is available as well.
Importantly, to simulate the high-frequency rotation, which is essential for adiabatic charge pumping, $d\tau=10^{-6}s$ is adopted hereafter.
Supposing $[0,2\pi]$ is divided into $l$ intervals, one has $f \approx 1/(l d \tau)$.
The variation of $d\tau$ does not change the main results of this paper.
 Due to the existence of band gaps [see Fig. \ref{f1}(b)], $n_j$ for both bulk and edge states are insensitivity to the rotational angle $\theta$, and the corresponding $I_{j,L/R}$ can be neglected. We next only pay attention to the charge pumping of zero-energy modes, which is closely related to the corner states of the HOTIs.

\section{eliminating the influence of finite size effect}\label{section3}
Significantly, a constrain in our numerical method should be considered. If $\psi_j$ describes a corner state, such a corner state must be isolated.
Fortunately, this condition can be easily satisfied in realistic samples. The environment (e.g. disorder, the coupling between the leads, etc.) will induce the energy difference, and their values are much larger than the coupling strength between two corner states.
Nevertheless, in order to simulate a faithful charge pumping process, one should eliminate the finite size effect in numerical calculation.
Due to the finite size effect, there exists a weak coupling between the zero-energy corner states, which can be dealt with by the degenerate perturbation.
Such a degenerate perturbation dramatically changes the definition of the $ n_{j}= \langle b_{j}^{\dagger}b_{j}\rangle$ and significantly influences the charge pumping as follows.

\begin{figure}[t]
	\centering
	\includegraphics[width=0.49\textwidth]{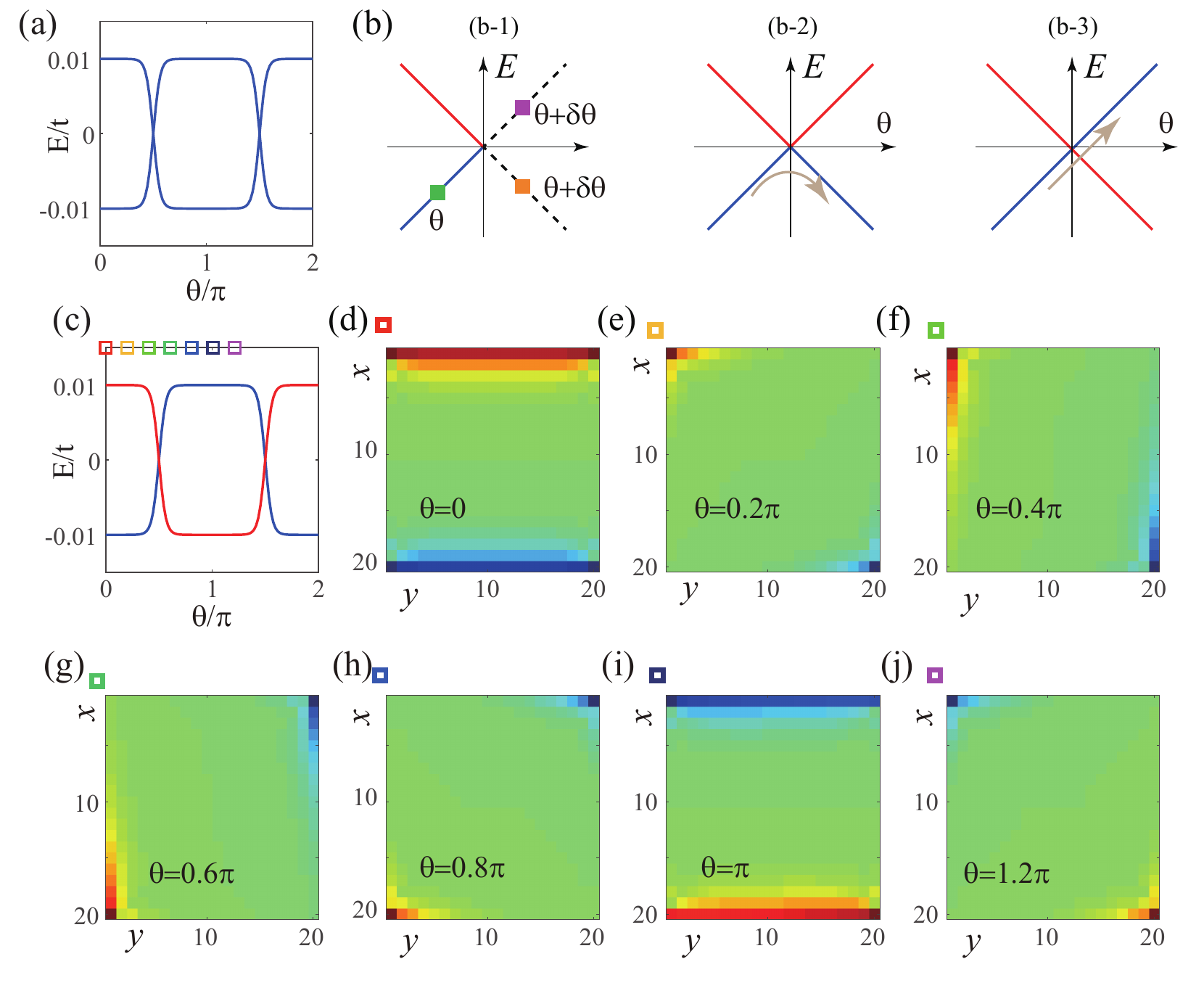}
	\caption{The voltage potential $H_{U}$ is introduced to eliminate the degenerate perturbation between two corners due to the finite size effect. $H_{U}$ satisfies Eq. (\ref{EQ9}). (a) The evolution of the corner states with $U=0.01t$. (b-1) shows the degenerate point of eigenvalues in (a). (b-2) and (b-3) are two possible evolution paths by increasing $\theta$. (c) is the obtained evolution path based on the adiabatic features of the pumping process. (d)-(j) are the wavefunctions for the red and blue lines in (c) by varying $\theta$. }
	\label{f2}
\end{figure}

 Taking $\theta=0.25\pi$ as an example, the zero-energy states for different corners can be marked as $ |1,1\rangle $ and $ |N,N\rangle $.
 After considering the coupling strength $\Delta$ induced by the finite size effect, the effective perturbation Hamiltonian for these zero-energy modes reads $[
 \begin{array}{cc}
 0 & \Delta\\
 \Delta &0\\
 \end{array}]$.
 For $ \Delta =0 $, its eigenfunctions are $ |\varphi_{1} \rangle = |1,1\rangle =[1,0]^T $ and $|\varphi_{2} \rangle = |N,N\rangle= [0,1]^T$, with eigenvalues $ \varepsilon_{1}=\varepsilon_{2}=0 $. $T$ stands for the transpose.
By contrast, for $ \Delta \neq 0 $ [no matter how weak $\Delta$ is], the degenerated zero energy states are lifted by the degenerate perturbation, and the eigenfunctions become $ |\varphi_{1,2} \rangle = \frac{1}{\sqrt{2}}[|1,1\rangle \pm |N,N\rangle] $.
 Then, a well defined $ n_{j}= \langle b_{j}^{\dagger}b_{j}\rangle$ for each corner state is not available.
 More importantly, the charge hopping between the sample and the left/right lead will happen non-locally when $\Delta\neq 0$.
 Such a nonlocal feature is improper since the transport between the leads and the sample occurs locally in the experiment.
 As a result, $\Delta$ and the degenerate perturbation should be eliminated to simulate the experimental setups.

To overcome such a problem, we introduce an additional perturbation in Eq. (\ref{EQ2}), where a step-potential along the $ x $ direction is adopted \cite{42}:
\begin{align}
	\begin{split}
		H_{U}=\sum_{n,m}2U[\frac{1}{2}-\Theta (\frac{N}{2}-n)]\tau_0\sigma_0 c_{n,m}^{\dagger}c_{n,m}.
\label{EQ9}
	\end{split}
\end{align}
$U$ represents the small voltage potential, and $ \Theta (N/2-n) $ is the step function.
 Figure \ref{f2}(a) plots the eigenvalues versus $ \theta $  for the corner states with $ U=0.01t$.

The second question is that the evolution of eigenvalues of the corner states versus $\theta$ intersects two times during one pumping cycle after considering Eq. (\ref{EQ9}) [see Fig. \ref{f2}(a)].
The evolution paths are ambiguous at the crossing points, where there are two possible cases [see Fig. \ref{f2}(b)].
Nevertheless, the charge pumping process should preserve the adiabatic features \cite{22} with the eigenfunctions evaluated continuously.
Based on the continuity condition of eigenfunctions, one is able to identify the correct evolution path by comparing the overlap of wavefunctions.
Specifically, we denote the eigenfunctions for the green, orange and purple dots in Fig. \ref{f2}(b-1) as $ \psi_{\theta,j} $, $ \psi_{\theta + \delta \theta,j} $, and $ \psi_{\theta + \delta \theta,j+1} $. Since $ |\langle \psi_{\theta,j}|\psi_{\theta + \delta \theta,j+1} \rangle|^{2} > |\langle \psi_{\theta,j}|\psi_{\theta + \delta \theta,j} \rangle|^{2} $ for $ \delta \theta \rightarrow 0 $, the eigenfunctions evolve along the path shown in Fig. \ref{f2}(b-3).

Now, we capture the reasonable eigenvalue and eigenvector evolutions for different corner states as shown in Figs. \ref{f2}(d)-(j). The colors of the plots, marked in red and blue, correspond to the red and blue eigenvalues shown in Fig. \ref{f2}(c), respectively.  It is obvious that the wavefunctions for different corners are completely decoupled.
Furthermore, the wavefunctions are always localized at the edge or corner rather than the bulk, which is unique for the HOTIs.
Based on the above results, we can study the charge pumping in realistic samples.

\section{ quantized charge pumping based on the mass domain walls of HOTI}\label{section4}

\begin{figure}[t]
	\centering
	\includegraphics[width=0.48\textwidth]{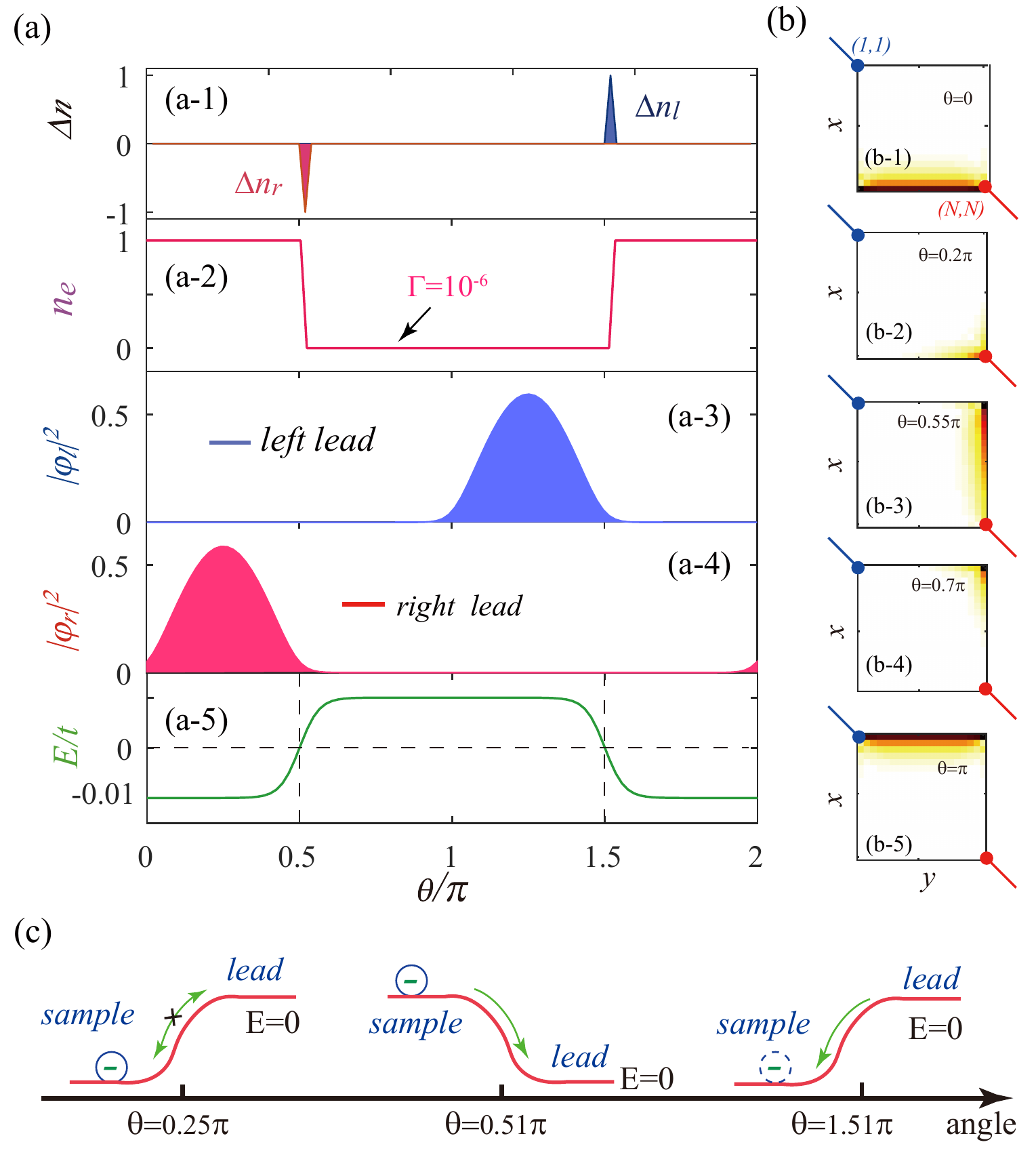}
	\caption{(a-1) is the evolution of the varied charge density $\Delta n_{r,l}$ versus $\theta=2\pi f\tau$ for the right lead (red) and left lead (blue), respectively. $f=10^{4}$ is adopted without specific statement.
(a-2) The evolution of the occupation number $n_e$ versus $\theta$.
(a-3) and (a-4) are the eigenstates $|\varphi_l|^2$ and $|\varphi_r|^2$ for the eigenenergy shown in (a-5), respectively.
(a-5) is the eigenenergy versus $\theta$. Since the two curves in Fig. \ref{f2}(e) can be analyzed in a similar manner, we only pay attention to one of the curves.
The corresponding wavefunctions for different $\theta$ are shown in (b-1)-(b-5). The blue and red marks denote the left and right leads and the contact points. The parameters are $U=0.01t$, $E_F=0$ and $\Gamma=10^{-6}t$.
(c) shows three schematic diagrams of charge transfer processes for typical $\theta$ in (a). The solid (dashed) circle indicates that the energy level is filled (empty) with $n_e=1$ ($n_e=0$). The red solid lines mark the energy for the sample and the lead. From left to right: electron transfer is forbidden; from sample to lead; from lead to sample. }
	\label{f3}
\end{figure}

As presented in the previous section, the domain wall-protected boundary states rotate with magnetic field.
When a lead is coupled with the sample, the charges carried by the bound states will jump into the lead or vice versa, and the charge pumping current becomes available.
In this section, we present the picture of quantized charge pumping based on the corner states of HOTIs.
Figure \ref{f3}(b-1) displays the schematic diagram of the charge pumping setup.
The leads' Fermi energy is fixed to $ E_{F}=0 $ with $V_{L/R}=0$, and the linewidth function is set as $ \Gamma_{L}=\Gamma_{R}=\Gamma=10^{-6}t $ \cite{41} unless otherwise noted.

Before proceeding, we have a few remarks based on Eq. (\ref{EQ7}), which are important to clarify the charge pumping process.
 (i) The pumping current $ I_{L/R} $ and the occupation number $ n_{j} $ dramatically depend on $ \widetilde{\Gamma}_{L/R}=\Gamma_{L/R}|\psi_{l/r}|^{2} $. Here, $ \psi_{l}=\psi_{j}(1,1) $ and $ \psi_{r}=\psi_{j}(N,N) $.
 This implies that the corresponding $ \widetilde{\Gamma}_{L/R} $ is non-zero if $\psi_{l,r}\neq0$.
 For $ \widetilde{\Gamma}_{L/R} \neq0$, the sample can exchange charges with the leads.
 (ii) Aside from $ \widetilde{\Gamma}_{L/R} $, the variation of $ n_{j} $ also relies on the relative values of the eigenvalue $ E_{\theta,j} $ for the sample and the Fermi energy $ E_{F} $ for the leads, as shown in Fig. \ref{f3}(c).
 When the leads are coupled with the sample's eigenvectors with $ \psi_{l/r} \neq 0 $, electrons will transfer from the lead to the sample in the case of $ E_{\theta,j}<E_{F} $. Alternatively, electrons will be transferred from the sample to the lead when $ E_{\theta,j}>E_{F} $.

Now, we demonstrate the pumping process by manipulating the bound states of the domain wall-protected HOTIs.
For simplicity, we only pay attention to one of the corner states [$j=2N^2$ for $\theta=0$] since they are decoupled, and the rest case can be analyzed in a similar manner.
We display the evolution of several  typical quantities in Figs. \ref{f3}(a-1)-(a-5).
It plots the wavefunctions coupled to the left/right lead $|\varphi_{r/l} |^2 \equiv|\psi_{\theta,j; r/l}|^{2} $; the corresponding eigenenergy $ E\equiv E_{\theta,j} $ and the occupation number $ n_e\equiv n_{j} $. Here, $n_j$ is calculated numerically by solving Eq. (\ref{EQ7}) self-consistently.

At the beginning of the pumping with $\theta=0$, states with energy $ E $ smaller than $ E_{F} $  are occupied (i.e., $ n_{e}=1 $), and the wavefunction $ |\varphi_{r,l}|^2 $ is concentrated at the lower boundary [see Figs. \ref{f3}(b) and (c)].
Although the eigenstates are coupled with the right leads [$\widetilde{\Gamma}_{R}\neq0$], charge exchange is not available since the state is fully occupied with $n_e=1$ and $ E<E_F $.
With the increasing of $\theta$, the corner state localized at $ (N,N) $ emerges and correspondingly $ |\varphi_{r}|^2 $ reaches its maximum.
By further increasing $ \theta $, the wavefunction gradually extends along the $x$ direction and moves toward the higher energy.
Notably, for the critical $ \theta=\pi/2 $ where $ E = E_{F} $, the wavefunctions are still coupled to the right lead with $ |\varphi_{r}|^2\neq0 $ [see Figs. \ref{f3}(a-3) and (b-3)].
Therefore, the charge will be transferred from the sample to the right lead [see Fig. \ref{f3}(c)] when $\theta>\pi/2$, which corresponds to the discharging process with $\Delta n_r\equiv dn_{j,R}=-1$.
Then, the state is empty, and $ n_{e} $ drops to zero [see Fig. \ref{f3}(a-2)].

\begin{table}[t]
\centering
\caption{ Several typical charge exchange processes for different $\theta$ in Fig. \ref{f3}(a). $E$ is the energy plotted in Fig. \ref{f3}(a-5). $E_F$ is the Fermi energy for the leads. The first row shows the $\theta$ for the corresponding quantities. For instance, $n_e$ is the occupation number slightly before $\theta$ [ i.e., $\theta-\delta\theta$]. }\label{TAB2}
\setlength{\tabcolsep}{1.3mm}{
\begin{tabular}{cccc|c}
\hline
\hline
$\theta$ & $\theta$ & $\theta$ & $\theta-\delta\theta$ & events at $\theta$\\
\hline
$\varphi_r\neq 0$ & $\varphi_l= 0$ & $E<E_F$ & $n_e=1$ & {\it no charge exchange}\\
$\varphi_r\neq 0$ & $\varphi_l= 0$ & $E>E_F$ & $n_e=1$ & {\it discharge at right lead}\\
$\varphi_r=0$ & $\varphi_l= 0$ & $E>E_F$ & $n_e=0$ & {\it no charge exchange}\\
$\varphi_r= 0$ & $\varphi_l\neq 0$ & $E>E_F$ & $n_e=0$ & {\it no charge exchange}\\
$\varphi_r= 0$ & $\varphi_l\neq 0$ & $E<E_F$ & $n_e=0$ & {\it charge at left lead}\\
$\varphi_r= 0$ & $\varphi_l= 0$ & $E<E_F$ & $n_e=1$ & {\it no charge exchange}\\
\hline
\hline
\end{tabular}}
\label{TABLE1}
\end{table}

A similar analysis can be applied for the rest half of the cycle, and the charging process is achieved at the left lead.
In simple terms, the whole pumping process is to discharge at the right and charge at the left lead, as summarized in TABLE. \ref{TABLE1}. Consequently, one electron is transferred from the left lead to the right one during each cycle. Because of the existence of two equivalent energy levels, the system pumps a total of two electrons, which is quantized in one period.
We have to emphasize that the charge is carried by the corner states and do not pass through the bulk during the pumping process, which is unique for HOTIs.

\section{manipulating the pumping current}\label{section5}

Having established the pictures of charge pumping process in HOTIs, we investigate the effect of external parameters on the pumping current in this section.
Since the conservation of current, i.e., one has $I_L=-I_R$, we only focus on $I_L$ in the following.

\begin{figure}[t]
	\centering
	\includegraphics[width=0.48\textwidth]{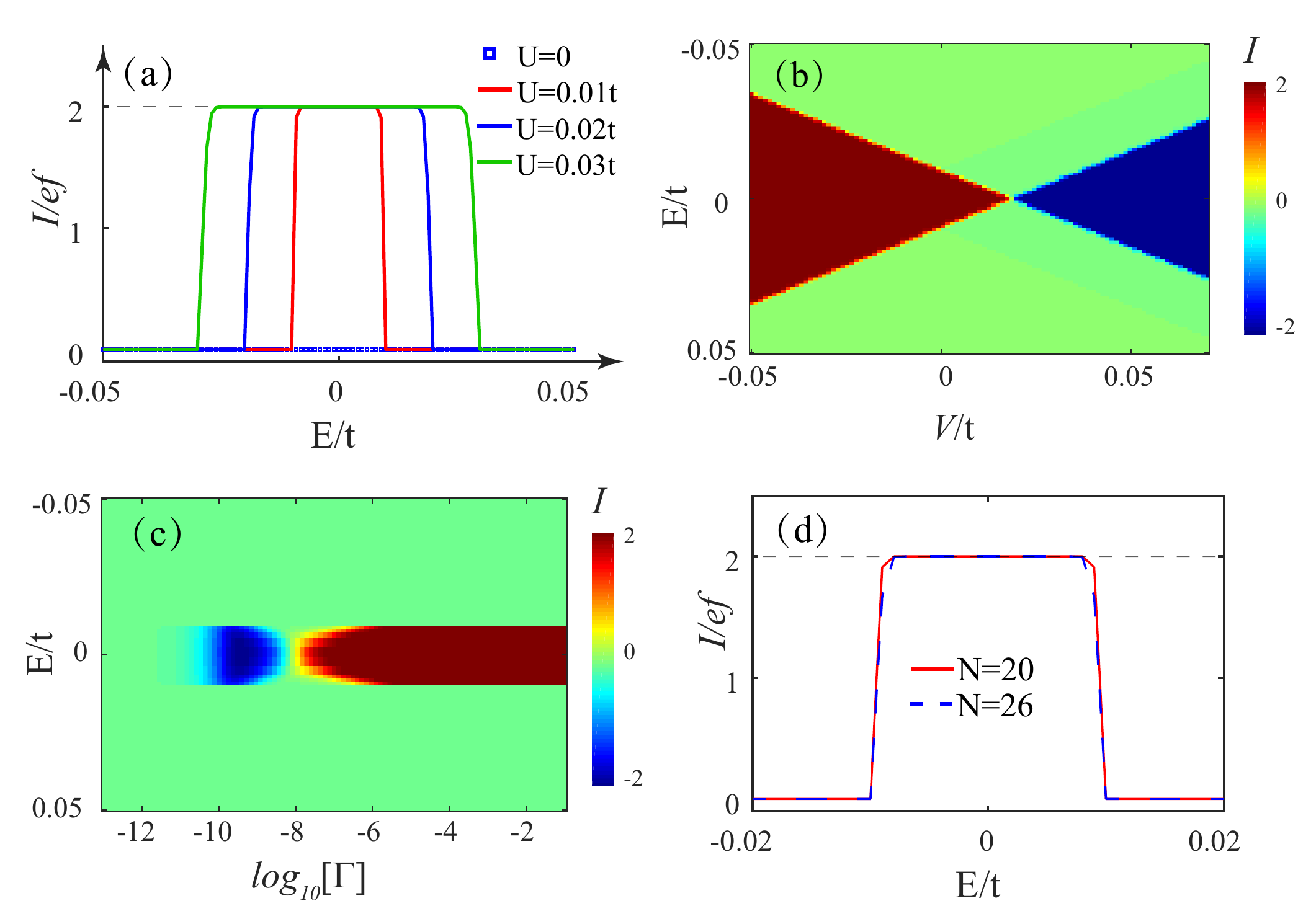}
	\caption{(a) Pumping current $I=I_L$ versus Fermi energy $E$ for different $U$. (b) Pumping current $I$ versus $(E,V)$ with $\pm V/2$ the additional voltage potential on left ($-V/2$) and right leads ($V/2$) with $V_{L/R}=\pm V/2$. (c) $I$ versus $E$ and $\log_{10}[\Gamma]$. $\Gamma$ is the linewidth function for leads, which denotes the coupling strength between the leads and the sample. (d) The sample size dependence of $I$ versus $E$.  }
	\label{f4}
\end{figure}

As illustrated in Fig. \ref{f4}(a), the current $I_L$ in the left lead is zero when the voltage potential $ U=0 $ because the degenerate perturbation between different corners does not allow the charging and discharging processes to occur independently.
As clarified in Sec. \ref{section3}, such a case cannot happen in realistic materials.
However, once the voltage potential $ U $ is introduced, the charge pumping appears and gives rise to plateaus of quantized currents, even for $ U=0.01 $.
By further increasing $ U $, the quantized current plateau becomes wider because the energy range of the $E-\theta$ curve broadens, as shown in Fig. \ref{f3}(a-5).
Besides, the current $ I$ is almost unchanged by increasing $N$ [see Fig. \ref{f4}(d)], manifesting that the quantized charge pumping is insensitive to the sample size $N$.

When the voltage potentials $ V_{L}=- \frac{V}{2} $ and $V_{R}=+\frac{V}{2} $ for the leads are considered, the effective Fermi energy for the leads are modified to $E_F+V_{L/R}$.
As shown in Fig. \ref{f4}(b), $ V $ can adjust the quantized pumped current $I$ and can even reverse the sign of $I$.
 Compared with $ V=0 $ in Fig. \ref{f3}(a), $ V $ changes the relative energy between the leads and the sample, thus affecting the charging and discharging processes.
 For $V<0$, the jump point of $n_j$ shifts to smaller (larger) $\theta$ [lower (higher) energy] for the right (left) lead, which maintains the direction of the pumping current.
While for $V>0$, the jump point of $n_j$ moves toward larger (smaller) $\theta$ for the right (left) lead.
 When $E_F-|V_{L}|<|U|$ and $E_F+|V_{R}|>|U|$, the direction of the pumping current should be reversed since the discharging process in the right lead is unavailable.

\begin{figure}[t]
	\centering
	\includegraphics[width=0.49\textwidth]{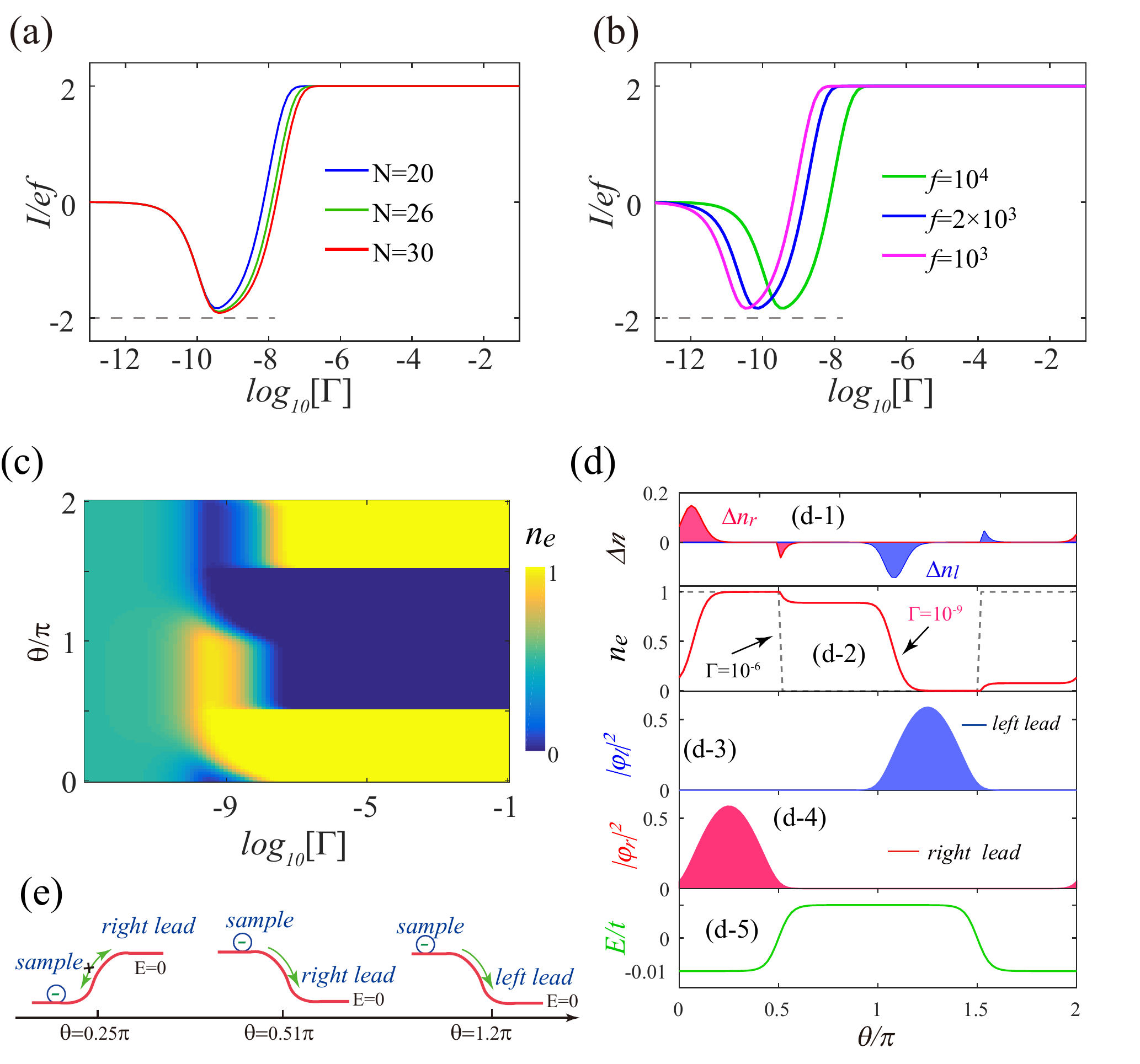}
	\caption{(a) Pumping current $I$ versus $\log_{10}[\Gamma]$ for different sample size $N$. (b) $I$ versus $\log_{10}[\Gamma]$ for the frequency $f$ with $\theta=2\pi f\tau$. (c) Occupation number $n_e$ for eigenvalue $E_{j=2N^2}$ versus angle $\theta$ and $\log_{10}[\Gamma]$. (d-1)-(d-5) The $ n_{e},\Delta n$ and $|\varphi_{l/r}|^2,E $ versus $ \theta $. The parameters are the same with Fig. \ref{f3} except $\Gamma=10^{-9}t$.
(e) shows three schematic diagrams of charge transfer process for different $\theta$ in (d). }
	\label{f5}
\end{figure}

 Interestingly, similar phenomena can also be observed in Fig. \ref{f4}(c), where the coupling strength $ \Gamma $ for the leads can also adjust the magnitude and the direction of $I$.
 The reversal of $I$ induced by $ \Gamma $ is seldom reported in previous studies and will be analyzed in detail below.

First, we need to rule out the influence of finite size effect and rotational frequency $ f $.
Figure \ref{f5}(a) shows the pumping current $ I $ versus $ \log_{10}[\Gamma] $ for different sample size $N$. As $ N $ increases, the reversal persists, demonstrating that the reversal is not attributed to the finite size effect.
Importantly, since the wavefunctions of the bound states become more localized for a larger sample size, $ \widetilde{\Gamma}_{L/R}=\Gamma|\psi_{l/r}|^{2} $ decreases in this case.
Thus, the critical point, where $I$ changes its sign, moves towards higher $ \Gamma $.
Similarly, the variation of rotational frequency $ f $ does not affect the current reversal either [see Fig. \ref{f5}(b)].
Nevertheless, the lower the frequent $ f $ is, the easier it is for leads to exchange charges with the sample.
Namely, decreasing $ f $ is equivalent to enhancing $ \Gamma $, and the critical point shifts toward the lower $ \Gamma $ as $ f $ decreases [see Fig. \ref{f5}(b)].

According to the above results, the reversal of $I$ only depends on the $ \Gamma $ and it is a physically reliable process.
We then investigate the occupation number $ n_{e} $ versus $ (\theta,~\log_{10}[\Gamma]) $.
Comparing Figs. \ref{f5}(a) and (c), it is clear that $I$ deviates from the quantized value when the occupation number $ n_{e} $ versus $ \theta$ significantly changes.
Besides, $ n_{e} $ changes its value as $ \Gamma $ decreases in some regions.
The state which is originally occupied [$n_e=1$] becomes empty [$n_e=0$], and the empty state is occupied.
Accordingly, the current is reversed since the occupation numbers are approximately opposite for large and small $\Gamma$.

For the sake of clarity, we plot $ n_{e},\Delta n$ and $|\varphi_{l/r}|^2,E $ in Figs. \ref{f5}(d-1)-(d-5) in the case of weak coupling $ \Gamma=10^{-9}t $, which accounts for the current reversal.
We start from the point with $ n_{e}=1 $ such as $ \theta =0.25\pi $.
As $\theta$ exceeds $0.5\pi$, the energy of the bound state $ E $ surpasses the leads' Fermi energy $ E_{F} $.
Therefore, the occupation number $ n_{e} $ starts to decrease, which seems to be the same as the case of $ \Gamma=10^{-6}t $.
Nevertheless, unlike the case with larger $\Gamma$ [see Fig. \ref{f5}(d-3)], $n_e$ does not directly reduce from one to zero.
The occupation number only changes slightly with $|\Delta n|\ll 1$ [see Figs. \ref{f5}(d-1) and (d-2)].
Thus, an almost fully filled state preserves when it leaves the right lead although $E>E_F$ is achieved.

Continuing to increase $ \theta $, the remaining electrons should be released in the left lead since $E>E_F$ still holds.
Because of the sufficiently long discharge time in the left lead, almost all the charges are released in the left lead [see Figs. \ref{f5}(d-2) and (c)].
Similarly, an almost empty state leaves the left lead with the charging process negligible.
Thus, the charge pumping processes in the left and right leads are opposite to those in Fig. \ref{f3}(a), which induces the reversal of the pumping current. In short, the reversal of the pumping current arises from the insufficient discharge and charge processes in the right and left leads, respectively.

\section{conclution}\label{section6}	

In summary, the quantized charge pumping in HOTIs is studied.
We found an interesting charge pumping process based on the unique topological natures of the HOTIs, which is distinct from the widely studied ones.
The charges only shift along the sample's boundary, and the bulk states do not directly participate in the pumping process.
Significantly, the bulk states ensure the existence of gapped edge states, so the high-order topological order is essentially required for the pumping process.
Furthermore, the manipulations of the pumping current by external parameters are also uncovered.
We find that the direction of the pumping current strongly depends on the coupling strength and the chemical potential of the leads.
Our work extends the understanding of exotic transport properties in HOTIs.

\section{Acknowledgements}
 We are grateful to Qiang Wei, Hongfang Liu, Rui-Chun Xiao, Chui-Zhen Chen and especially Qing-feng Sun for fruitful discussions. This work was supported by the National Basic Research Program of China (Grant No. 2019YFA0308403), NSFC under Grant No. 11822407 and No. 12147126. A.M.G. acknowledges supports from the NSFC under Grant No. 11874428 and the High Performance Computing Center of Central South University .

\end{document}